\documentclass[prl,twocolumn,nofootinbib, preprintnumbers, superscriptaddress]{revtex4}

\usepackage{amsmath,amssymb,slashed,braket}
\usepackage{graphicx}
\usepackage{epstopdf}
\usepackage{float}
\usepackage[colorlinks=true,
            linkcolor=blue,
            urlcolor=blue,
            citecolor=green,          
            bookmarks=true,
            bookmarksnumbered=true,
            breaklinks=true,
            pdfpagemode=Fullscreen,
            pdfstartview=FitBH]{hyperref}

\usepackage[normalem]{ulem}
\usepackage{color}

\definecolor{Orange}{cmyk}{0,0.61,0.87,0}
\definecolor{JungleGreen}{cmyk}{0.99,0,0.52,0}
\definecolor{OliveGreen}{cmyk}{0.64,0,0.95,0.40}
\definecolor{Brown}{cmyk}{0,0.81,1,0.60}
\definecolor{RoyalBlue}{cmyk}{0.71,0.53,0,0.12}
\definecolor{Gray}{cmyk}{0,0,0,0.40}
\definecolor{LightPink}{cmyk}{0.0,0.25,0,0}
\definecolor{LLightPink}{cmyk}{0.0,0.10,0,0}
\definecolor{LightBlue}{cmyk}{0.25,0,0,0}
\definecolor{LightGray}{cmyk}{0,0,0,0.2}

\usepackage{xcolor}
\definecolor{gesfpurple}{rgb}{0.47,0.19,0.42}

\definecolor{gesflanse}{rgb}{0.00,0.50,0.50}

\definecolor{gesfblue}{rgb}{0.08,0.42,0.76}

\definecolor{gesfred}{rgb}{1,0,0}

\definecolor{gesfwhite}{rgb}{1,1,1}

\definecolor{gesfblack}{rgb}{0,0,0}

\newcommand{\geqn}[1]{Eq.\,\hypersetup{linkcolor=blue}(\ref{#1})\hypersetup{linkcolor=blue}}
\newcommand{\gfig}[1]{{\hypersetup{linkcolor=violet}Fig.\,\ref{#1}\hypersetup{linkcolor=blue}}}

\graphicspath{{figs/}}

\begin{document}

\title{Right-Handed Neutrino Dark Matter with Forbidden Annihilation}

\author{Yu Cheng}
\email[Corresponding Author: ]{chengyu@sjtu.edu.cn}
\affiliation{Tsung-Dao Lee Institute \& School of Physics and Astronomy, Shanghai Jiao Tong University, China}
\affiliation{Key Laboratory for Particle Astrophysics and Cosmology (MOE) \& Shanghai Key Laboratory for Particle Physics and Cosmology, Shanghai Jiao Tong University, Shanghai 200240, China}

\author{Shao-Feng Ge}
\email{gesf@sjtu.edu.cn}
\affiliation{Tsung-Dao Lee Institute \& School of Physics and Astronomy, Shanghai Jiao Tong University, China}
\affiliation{Key Laboratory for Particle Astrophysics and Cosmology (MOE) \& Shanghai Key Laboratory for Particle Physics and Cosmology, Shanghai Jiao Tong University, Shanghai 200240, China}

\author{Jie Sheng}
\email[Corresponding Author: ]{shengjie04@sjtu.edu.cn}
\affiliation{Tsung-Dao Lee Institute \& School of Physics and Astronomy, Shanghai Jiao Tong University, China}
\affiliation{Key Laboratory for Particle Astrophysics and Cosmology (MOE) \& Shanghai Key Laboratory for Particle Physics and Cosmology, Shanghai Jiao Tong University, Shanghai 200240, China}

\author{Tsutomu T. Yanagida}
\email{tsutomu.yanagida@sjtu.edu.cn}
\affiliation{Tsung-Dao Lee Institute \& School of Physics and Astronomy, Shanghai Jiao Tong University, China}
\affiliation{Key Laboratory for Particle Astrophysics and Cosmology (MOE) \& Shanghai Key Laboratory for Particle Physics and Cosmology, Shanghai Jiao Tong University, Shanghai 200240, China}

\begin{abstract}
The seesaw mechanism with three right-handed neutrinos
has one as a well-motivated dark matter candidate if
stable and the other two can explain baryon asymmetry via
the thermal leptogenesis scenario. We explore the
possibility of introducing additional particles to make
the right-handed neutrino dark matter in thermal equilibrium
and freeze out through a forbidden annihilation channel.
Nowadays in the Universe, this forbidden channel
can be reactivated by a strong gravitational potential
such as the supermassive black hole in our galaxy center. 
The Fermi-LAT gamma ray data and dark matter relic density
require this right-handed neutrino dark matter to have mass
below $100\,$GeV and the existence of an additional boson $\phi$
that can be tested at future lepton colliders.

\end{abstract}

\maketitle 

{\bf Introduction} -- The heavy right-handed Majorana neutrinos are widely
considered as a key ingredient beyond the Standard
Model of particle physics. They explain not only the
observed tiny neutrino masses via the seesaw mechanism
\cite{Minkowski:1977sc,Yanagida:1979as,Yanagida:1979gs,Gell-Mann:1979vob}, but also the Universe's baryon asymmetry
through leptogenesis \cite{Fukugita:1986hr}. Normally, we assume three
heavy right-handed neutrinos (RHN). However, two heavy
Majorana neutrinos are sufficient to explain the
observed neutrino oscillation data and the baryon asymmetry in our Universe \cite{Frampton:2002qc,Raidal:2002xf,Glashow:2003nk,Barger:2003gt,Ge:2010js}.
One remaining right-handed neutrino can be either very heavy or very
light \cite{Kusenko:2010ik,Drewes:2013gca,Merle:2013gea,Drewes:2016upu,Dasgupta:2021ies}. In this letter, we explore the latter option that
provides us with interesting consequences at low energies.

If the remaining right-handed neutrino is as light as
$\mathcal O$(100)\,GeV, it can be a good candidate of
dark matter (DM) \cite{Kusenko:2009up,Merle:2013gea,Drewes:2016upu,Abazajian:2017tcc,Boyarsky:2018tvu,Dasgupta:2021ies}. 
However, we have strong constraints from cosmic microwave background (CMB)
if they annihilate into the SM particles and the subsequent
electromagnetic energy injection affects the ionization history of
the universe \cite{Bean:2003kd,Chen:2003gz,Hansen:2003yj,Pierpaoli:2003rz,Padmanabhan:2005es,Slatyer:2009yq,Roszkowski:2017nbc,Planck:2018nkj,Cang:2020exa,Kawasaki:2021etm,Liu:2023nct}.
Therefore, we consider a forbidden-type DM \cite{DAgnolo:2015ujb,Delgado:2016umt,DAgnolo:2020mpt,Wojcik:2021xki}
whose freeze-out production is through the annihilation channel
to a heavier dark sector partner or SM particles.
In other words, DM annihilation is kinematically forbidden.
The only way out is the thermal energy that can overcome the mass
difference between the initial and final states. This can happen
at the early Universe but ceases when the temperature cools down
with the Universe expansion. Around the last scattering
of CMB photons, the temperature has dropped to eV scale which is
negligibly small for any reasonable forbidden DM model.
So forbidden DM can avoid those CMB constraints.

Without annihilation at present time, it is very difficult
for observation to verify the existence of this forbidden scenario.
The indirect detection is intrinsically forbidden.
Interestingly, a strong gravitational source, 
such as a supermassive Black Hole (SMBH), 
can reactivate the forbidden DM and make the 
DM annihilate around it \cite{Cheng:2022esn}.  
The subsequent decay of forbidden channel final states
into SM particles, especially visible photons, can give
a unique indirect detection signature today. The gamma
ray from the DM forbidden annihilation only appears
in the vicinity around the SMBH, but not anywhere else in
the sky. A naive fit with the Fermi-LAT data for Sgr A*
shows quite good sensitivity on the thermally averaged
annihilation cross section there.

Although the Fermi-LAT constraint is consistent with
the DM relic density requirement with naive assignments,
a concrete model is necessary to demonstrate the realistic
possibility of constructing such a forbidden DM theory.
In this paper, we show our right-handed neutrino DM model
and its parameter space to generate the correct DM
relic density and at the same time provide reactivated
annihilation signal.
Comparing with the Fermi-LAT data, we find the 
right-handed neutrino DM mass is below $100$\,GeV and the presence
of an additional boson $\phi$ is required in the almost same
mass region. Finally, we point out that the possible
test at future lepton collider.

{\bf The Right-Handed Neutrino Forbidden DM} --
We assume a $Z_2$ parity acting only on the right-handed
neutrino $N$ to decouple it from all the
SM particles. Then, $N$ becomes stable and a good
DM candidate \cite{Cox:2017rgn} 
(If it is the case, we have an anthropic reason for the presence of three families of quarks and leptons \cite{Ibe:2016yfo}.) However, its production
is uncertain since it decouples completely from the SM
particles. It may be produced in the early Universe through
interactions with unknown heavy particles, but its abundance
is undetermined by low energy physics.

A possible solution is to introduce a fermion $\chi$ and a boson
$\phi$ to couple with the DM $N$, $\phi N \chi$. Notice here that
all fermions, $N$ and $\chi$, are left-handed two component
Weyl fermions. The DM neutrino $N$ is
assumed to carry odd parity under the $Z_2$. If $\phi$ is even (and $\chi$ is odd),
it can couple to a pair of the SM Higgs bosons, $H$ and $H^{\dagger}$, via $\phi H^{\dagger} H$.
Thus, the interaction Lagrangian is, 
\begin{equation}
  \mathcal L_{\rm int}
=
  (y \phi N \chi + h.c.)
+ \lambda m_{\phi} \phi H^{\dagger} H,
\label{Interaction}
\end{equation}
where $m_\phi$ is the mass of $\phi$.
The coupling $\lambda$ with the SM Higgs boson renders $\phi$ in
thermal equilibrium with the SM particles in the early Universe.

To maintain equilibrium, the decay rate of $\phi$ to SM fermions, 
$\Gamma \sim \frac 1 {32 \pi} \lambda^2 m^2_f m^3_\phi/(m_h^2 - m_\phi^2)^2$, 
should be larger than the Hubble constant $H \propto T_f^2/M_{\rm pl}$
where $M_{\rm Pl}$ is the Planck mass.
Suppose the $\phi$ decoupling happens around $T_f \sim m_\phi / 25$,
thermal equilibrium requires $\lambda \gtrsim (10^{-8} \sim 10^{-7})$
depending on the scalar mass $m_\phi$. Even lower limit is possible
if $m_\phi$ approaches the Higgs mass to reduce the denominator
$m^2_\phi - m^2_h$ of the decay width formula.
Further through the Yukawa term and the resulting
$N N \leftrightarrow \phi \phi$ scatterings, $N$
can also get in the thermal equilibrium.

All these new particles are singlets of the SM gauge
group and have mass terms,
\begin{equation}
  \mathcal L
=
  \frac{m_N}{2} NN
+ \frac{m_\chi}{2}\chi \chi
+ \frac{m_{\phi}^2}{2}\phi^2.
\label{Mass}
\end{equation}
For a forbidden scenario, the DM mass should be
lighter than its annihilation final-state particles,
$m_N < m_\phi$. Otherwise, the RHN $N$ can pairly
annihilate into a pair of $\phi$ at the late time and then
$\phi$ decays to SM particles through the mixing with the
SM Higgs boson.
For $\phi$ mass larger than $\mathcal O (1)$\,GeV,
the dominated decay channel is 
$\phi \rightarrow f \bar{f}$, 
where $f = b, \tau$. 
If $\phi$ mass is $\lesssim \mathcal O(1)$\,GeV,
$\phi$ can still decay through the di-photon channel
$\phi \rightarrow \gamma \gamma$
that is induced by triangle diagrams in the
similar way as the SM Higgs boson \cite{Workman:2022ynf}.
These processes are strongly constrained
by the CMB \cite{Elor:2015bho} and high energy
gamma ray data \cite{Fortes:2022cxp}. 
Therefore, we concentrate our discussion on the case
that the boson $\phi$ is heavier than the DM $N$,
$m_\phi > m_N$. 
Although the DM $N$ may annihilate into
SM fermions by exchanging a virtual Higgs,
$NN \rightarrow \phi + h^*$ and $h^* \rightarrow f \bar f$,
this three-body final state is naturally suppressed
by a phase factor $1/16 \pi^2 \sim 10^{-2}$ and 
the Yukawa coupling between Higgs and fermions
$y^2 \lesssim 10^{-3}$. In total, it has a
$\mathcal O(10^{-5})$ suppression relative to
the two body annihilation so that the CMB constraint
can hardly apply. We also consider 
that the mass of $\chi$ is sufficiently heavy,
$m_\chi > m_\phi + m_N$, such that $\chi$ can
decay and does not contribute to the DM relic
density. Below we focus on the density evolution of
the right-handed neutrino forbidden DM $N$.

In the beginning, the DM $N$ is also in thermal 
equilibrium. It gradually freezes out through the
forbidden annihilation channel
$N + N \leftrightarrow \phi + \phi$ 
by exchanging a $t$-channel $\chi$.
The number density evolution is governed by the 
Boltzmann equation,
\begin{equation}
\dot{n}_N+3 H n_N
=
- \left\langle\sigma_{N N} v\right\rangle
\left[
  n_N^2
- \left( n^{\rm eq}_N \right)^2
\right],
\label{Boltzmann}
\end{equation}
where $n_N$ ($n_N^{\rm eq})$ is the number density
of right-handed neutrino $N$ (in equilibrium).
Since the kinematic threshold of the forbidden
channel cuts the allowed range of the center-of-mass energy
$\sqrt s$, the total cross section cannot easily find an
analytical formula. It is more convenient to evaluate
the cross section $\langle \sigma_{\phi \phi} v \rangle$
of the reversed process $\phi \phi \rightarrow N N$
first and express
$\left\langle\sigma_{N N} v\right\rangle$ according
to detailed balance \cite{DAgnolo:2015ujb},
\begin{equation}
  \left\langle\sigma_{N N} v\right\rangle
\equiv
  \left( \frac{n_{\phi}^{\rm eq}}{n_N^{\rm eq}}\right)^2
  \left\langle\sigma_{\phi \phi} v\right\rangle
=
  \frac {(1+\delta)^3} 4 e^{-2 x \delta}
  \left\langle\sigma_{\phi \phi} v\right\rangle.
\label{InverseSigma}
\end{equation}
Being a function of the mass
difference $\delta \equiv (m_\phi - m_N)/ m_N$ 
and temperature parameter $x \equiv m_N/T$,
the exponential suppression $e^{-2 x \delta}$ 
comes from the 
ratio of number densities in equilibrium, 
$n_{\rm eq} = g(m T / 2 \pi)^{3 / 2} \exp (-m / T)$,
where $g=2$ and 1 for $N$ and $\phi$, respectively.
The Boltzmann equation can be solved semi-analytically
to obtain the rescaled number density
$Y_N \equiv n_N / T^3$,
\begin{equation}
    Y_N
=
    \frac{\left[ 
    (a-b \delta) \frac{1}{x_f} e^{-2 \delta x_f} g_{\delta}\left(x_f\right) 
+\frac{b}{2 x_f^2} e^{-2 \delta x_f}
    \right]^{-1}}
{
    \sqrt{\frac{\pi}{45}} 
        m_N M_{\mathrm{Pl}} (1 + \delta)^3
        g^s_* / \sqrt{g_*}
}
\label{eq:YN}
\end{equation}
where $x_f$ indicates the freeze-out temperature and
$g^s_*$ ($g_*$) the entropy (total) degrees of freedom then.
For convenience, we have defined
$g_{\delta}(x_f)$ $\equiv$ $1 - 2 \delta x_f$ $e^{2 \delta x_f}$
$\int_{2 \delta x_f}^{\infty}$ $t^{-1}$ $e^{-t}$ $d t$.
The parameters $a$ and $b$  are the expansion coefficients
of the thermally averaged cross section
$\left\langle\sigma_{\phi \phi} v\right\rangle$
for the $s$ and $p$ waves, 
$\left\langle\sigma_{\phi \phi} v\right\rangle
\equiv a + b x^{-1}$ \cite{Gondolo:1990dk}.
The DM relic density is related to $Y_N$ as,
\begin{equation}
  \rho_{\mathrm{DM}} = m_N s_0 Y_N,
\quad
  \Omega_{\mathrm{DM}} h^2
=
  \frac{\rho_{\mathrm{DM}}}{\rho_{c} / h^2},
\label{eq:rho}
\end{equation}
where $s_0 = 2891.2$ cm$^{-3}$ is today's entropy density and 
$\rho_c = 1.05 \times 10^{-5} h^2$\,GeV/cm$^3$ is the critical
density of the Universe.

\begin{figure}[!t]
\centering
\includegraphics[width=0.486\textwidth]{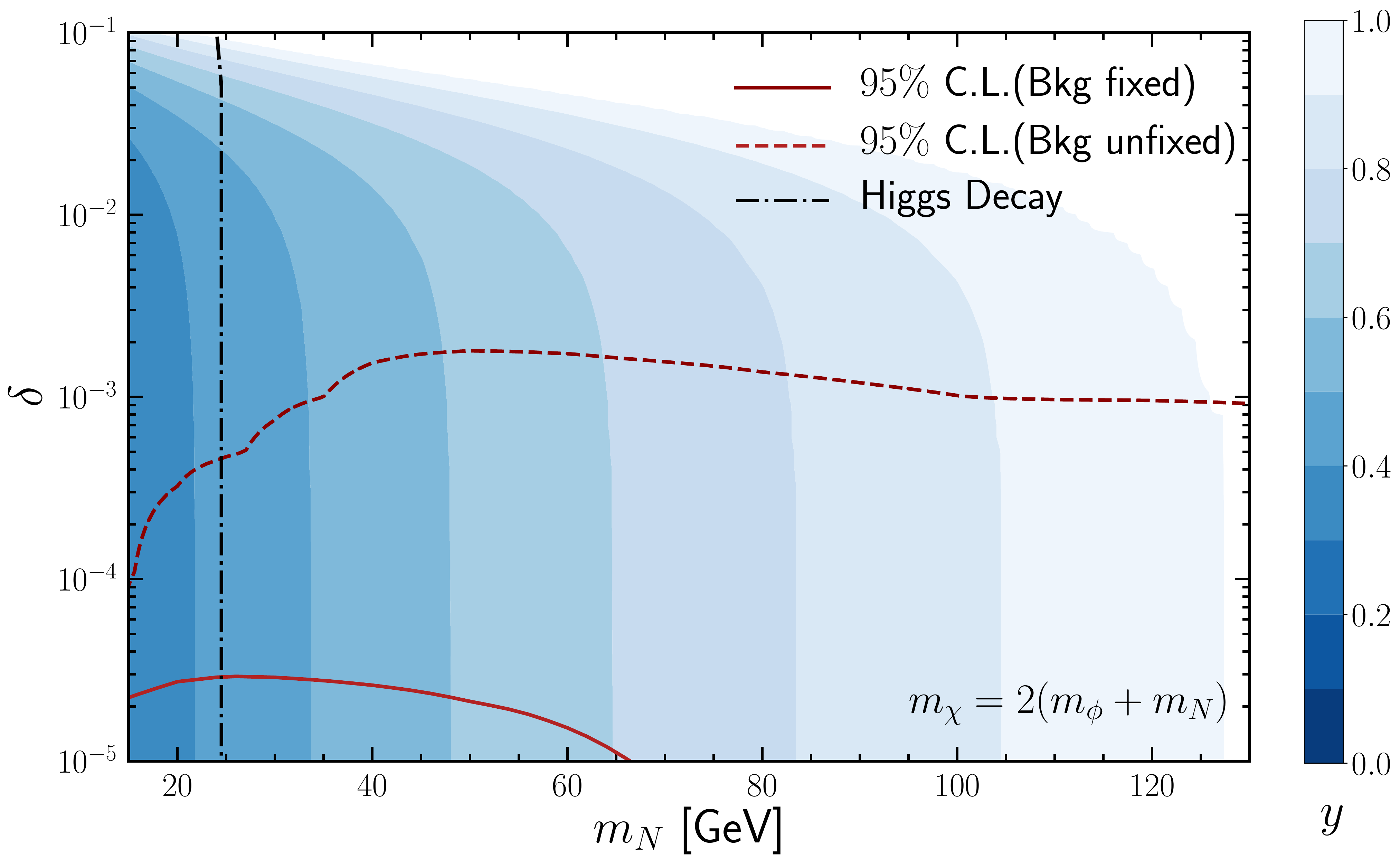}
\caption{
The allowed parameter space contour to obtain the DM relic
density through freeze out. After fixing non-DM fermion mass
$m_\chi = 2 ( m_N + m_\phi)$, there are still three model
parameters, the right-handed neutrino DM mass $m_N$ (horizontal
axis), the mass difference $\delta$ (vertical axis), and
the Yukawa coupling constant $y$ (contours according to the
color bar). Red lines show the 95\% limit provided by the
FermiLAT gamma ray data. Both cases with fixed (solid red)
and unfixed (dashed red) background are shown for comparison.
The almost vertical black dashed line indicates the 
maximial $m_N$ that allows Higgs invisible decay
$h \rightarrow N + \chi$.  
}
\label{Limit_Contour}
\end{figure}

We also solve the Boltzmann equation numerically with
{\tt MicrOMEGAs} \cite{Belanger:2020gnr}.
The allowed parameter space to produce the correct relic
density $\Omega_{\rm DM} h^2 = 0.12$ is shown in 
\gfig{Limit_Contour}. The two key mass parameters of the
forbidden channel are shown in terms of the right-handed
neutrino mass $m_N$ and the relative mass difference
$\delta$ as horizontal and vertical axes, respectively.
The whole $m_N$--$\delta$ parameter space is all allowed
by tuning the Yukawa coupling $y$ to match the DM relic
density. In other words, the value of $y$ is given as
a function of $m_N$, $\delta$, and $\Omega_{\rm DM}$.
 Although the partner fermion $\chi$ also has an
 adjustable mass $m_\chi$ that can enter the annihilation
 cross section, tuning its value affects
the DM relic density in the early Universe and the
reactivated forbidden annihilation at current days
simultaneously. This resembles the effect of
the Yukawa coupling $y$. Tuning the value of
either Yukawa coupling $y$ or $m_\chi$ is sufficient
for our purpose. For simplicity, we take fixed
$m_\chi = 2 (m_\phi + m_N)$ for illustration.
\gfig{Limit_Contour} indicates that
the DM relic density naturally limits its mass within the range of
$m_N \lesssim 100$\,GeV. A smaller coupling constant $y$
prefers smaller DM mass $m_N$ and mass difference $\delta$.
This is because when the annihilation
cross section becomes smaller, 
DM freezes out earlier and its 
number density becomes larger.
Since the DM
relic density is inversely proportional to the annihilation 
cross section $\braket{\sigma v}$, 
it scales with $m_N^2/y^4$. A smaller coupling
requires a lighter DM.
On the other hand, a smaller mass difference 
can increase the annihilation cross section to
compensate the effect of a small coupling.

{\bf The Forbidden DM Signature} --
The forbidden DM scenario has an intrinsic 
difficulty of being probed via indirect detection. 
Nowadays, the DM Boltzmann distribution with typical
$\mathcal O(100)$\,km/s velocity in our galaxy can
no longer support the forbidden channel. Such a
non-relativistic velocity $v \sim 10^{-3}$ can only
overcome the mass threshold $\delta \sim v^2 \sim 10^{-6}$.
Although direct detection and collider experiments
may provide complementary searches and even measure
the DM mass, the forbidden nature of DM is still missing.
In other words, the theoretical models of forbidden DM
can in principle exist but is not testable. This is a
very serious issue.

To reopen the forbidden channel, it is necessary to
accelerate the DM particles. Around a black hole,
especially the SMBH in the galaxy center, a particle 
can be accelerated to become relativistic because of the 
strong gravitational force \cite{Cheng:2022esn}.
This can help the DM $N$ to annihilate into heavier 
dark sector particle $\phi$ near the SMBH. The unstable
$\phi$ then decays to SM particles through its mixing with
the Higgs particle. The decay products are either gamma
photons or charged particles. While the di-photon decay
channel gives mono-energetic gamma rays in the $\phi$
rest frame, the charged particles produce photons through
final-state radiation or meson decays with continuous
spectrum. These allow using gamma ray 
that can point straightly back to the SMBH as signal
for testing our model of right-handed neutrino DM.
 
The gamma ray intensity highly depends on the DM
density and velocity dispersion profiles around the SMBH. 
In our scenario, DM $N$ has no self-interaction. Thus, 
its density follows the CDM density profile
\cite{Fields:2014pia}. For the innermost region,
$r < 4 G M \equiv r_0$ where $M$ is the BH mass, all particles are
attracted to fall into the SMBH so that the DM density is
$0$ therein. On the other hand, the DM halo simply
follows the NFW profile when the gravitational influence
of the SMBH no longer dominates which roughy happens at
$r_b \equiv 0.2 G M/v_0^2$ with $v_0$ being the DM
velocity dispersion there. In between, a DM density
spike forms,
\begin{equation}
\hspace{-2mm}
  \rho(r)
=
\begin{cases}
  0,
& r < r_0, \text { (Capture Region),}
\\
  \frac{\rho_{\mathrm{sp}}(r) \rho_{\mathrm{in}}(t, r)}{\rho_{\mathrm{sp}}(r)+\rho_{\mathrm{in}}(t, r)},
& r_0 \leq r < r_b,
  \text { (Spike), }
\\
  \text { NFW Profile}, 
&
  r > r_b,
  \text { (Halo). }
\label{CDM_Profile}
\end{cases}
\end{equation}
Naively thinking, the DM density profiles keeps increasing
when going towards the SMBH. However, the DM density
cannot increase forever. In addition to self-interaction,
annihilation can also suppresses the growth of DM density.
This naturally puts an upper bound on the DM density
profile and forms a density plateau around the SMBH.
Going inward from the spike region boundary $r_b$, the DM
density profile increases very fast initially as
$\rho_{\rm sp} (r) \equiv \rho_b (r_b/ r)^{\gamma_{\rm sp}}$
with $\gamma_{\rm sp} \equiv (9 - 2 \gamma_c)/ (4 -\gamma_c)$.
The density coefficient $\rho_b \equiv \rho_D (D/r_b)^{\gamma_c}$
with $\gamma_c = 1$ is scaled according to the NFW
profile \cite{Navarro:1995iw}, $\rho(r) \propto 1/r$ for $r \ll 26$\,kpc \cite{Alvarez:2020fyo,Abazajian:2020tww},
from the density around the Solar system with
$\rho_D = 0.3\,$ GeV cm$^{-3}$ and $D = 8.5$\,kpc. 
When reaching the SMBH vicinity, $r \gtrsim r_0$,
the DM density increases much slower as
$\rho_{\rm in} (r,t) \equiv \rho_{\rm ann} (t) (r/r_{\rm in})^{-\gamma_{\rm in}}$
with $\gamma_{\rm in} = 1/2$.
The density $\rho_{\rm ann}$ is the so-called annihilation plateau density 
$\rho_{\rm ann} \equiv m_N / \braket{\sigma_{N N} v} t$.
With $\rho = \rho_{\rm sp} \rho_{\rm in} / (\rho_{\rm sp} + \rho_{\rm in})$,
its value approaches $\rho_{\rm sp}$ for
$\rho_{\rm sp} \ll \rho_{\rm in}$ and
$\rho_{\rm in}$ in the opposite.
Since $\rho_{\rm sp}(r) \propto r^{-7/3}$
and $\rho_{\rm in}(r) \propto r^{-1/2}$,
the DM density profile almost becomes a plateau
in the inner spike region. The division $r_{\rm in}$
at $\rho_{\rm sp} = \rho_{\rm ann}$ is determined
by the DM annihilation cross section
$\langle \sigma_{N N} v \rangle$.

In its vicinity, the SMBH dominates the gravitational potential
and consequently the DM velocity dispersion follows
a simple scaling $v_d^2(r) \sim GM /r$.
In reality, the DM particles inside the spike region are
thermalized to a Juttner distribution
\cite{Juttner:1911,DeGroot:1980dk}
$P_r (V_r, V_c, x(r))$ where $V_r$ is the relative velocity
and $V_c$ is the centre-of-mass velocity of a two-particle
collision system \cite{Cheng:2022esn}. To get the energy
spectrum of the gamma production rate per volume
$d \Phi_\gamma (r) / d E_\gamma$, one needs to integrate
over the DM distribution from the threshold relative velocity $V^{\rm th}_r \equiv [1- 1/ (1-2 m_F^2/m_\chi^2)^2]^{1/2}$,
\begin{equation}
\hspace{-2mm}
 \frac{dF_{\gamma}}{dE_{\gamma}} (r)
=
  \int_{V^{\rm th}}^1 d V_{r} 
  P_r
  \left(V_{r},V_c, x(r)\right)
  \sigma V_{r}
  \frac{dN_{\gamma}}{dE_{\gamma}}(V_{r},V_c),
\label{eq:rate_spectrum}
\end{equation}
where $x(r)$ has radius dependence through the DM
temperature $T = \frac12 m_N v_d^2 (r)$ and
$dN_\gamma/dE_\gamma$ is the boosted photon spectrum
from $\phi$ decay. For the mass range $m_\phi > 10$\,GeV
considered in this paper, the dominate
channels are $\phi \rightarrow b \bar{b}$, $\tau \bar{\tau}$,
and $g g$. We use the {\tt PPPC4DMID} package
\cite{Cirelli:2010xx} to generate the photon spectra of
$\phi$ decay at rest and {\tt HDECAY} \cite{Djouadi:1997yw}
to obtain the branching ratios. 
After that, we boost the spectrum to the galaxy frame
\cite{Elor:2015bho,Elor:2015tva}.

The observed gamma ray flux is an integrated total result
over the radius from $r_0 = 4 G M$ to some upper limit $r_B$,
\begin{equation}
  \frac{d \Phi_{\gamma}}{d E_{\gamma}}
=
  \frac{1}{4 \pi D^2} \frac{1}{2 m_{\chi}^{2}} 
  \int_{4 G M}^{r_B} 4 \pi r^{2} d r  
  \rho^{2}(r) \frac{dF_{\gamma}}{dE_{\gamma}}(r).
\label{eq:gamma_flux}
\end{equation}
In principle, the integration upper limit $r_B$ should be
as large as possible to include all contributions. However, 
those DM particles at larger radius have a smaller
velocity dispersion and consequently only the tail of 
the Juttner distribution contributes. To make efficient
numerical evaluation, we only include those regions that,
\begin{equation}
  \int^1_{V^{\rm th}_r} d V_r \int_0^1 d V_c P_r(V_r, V_c, x(r_B))
\equiv
  1\%.
\label{eq:defrb}
\end{equation}
This means that only less than 1\% of the DM phase space
can contribute at radius larger than $r_B$.

{\bf Indirect Detection with Fermi-LAT} --
The photon flux from the forbidden annihilation 
is localized around the SMBH. Uniquely identifying
the right-handed neutrino DM scenario requires
not just observing the photon spectrum but also
good angular resolution. The Fermi-LAT satellite
observatory is an all-sky $\gamma$-ray telescope
with excellent energy and angular resolutions on the
$\gamma$-ray point source from the GC
\cite{Fermi-LAT:2009ihh,Fermi-LAT:2016uux}.
We analyze a square region of $10^\circ \times 10^\circ$
around Sgr\,$A^{*}$ and find several $\gamma$-ray point
sources, of which 4FGL\,J1745.6-2859 is the
brightest and closest one to Sgr\,$A^*$ in the
Fourth catalog of Fermi-LAT sources (4FGL)
\cite{Fermi-LAT:2019yla,Ballet:2020hze}.
This point source is usually considered as the
manifestation of Sgr\,$A^*$ in the
MeV-to-GeV range \cite{Cafardo:2021pqs}. We use 14
years of Fermi-LAT data from August 4, 2008 to October 26, 2022
in this paper.
To be specific, the Pass 8 SOURCE-class events from 100\,MeV
to 1000\,GeV are binned to a pixel size of $0.08^\circ$.

A universal model can describe the $\gamma$-ray spectrum
from point sources. To be specific, this universal
4FGL\,J1745.6-2859 spectral model is a log-parabola
in the 4FGL Catalog \cite{Fermi-LAT:2019yla,Ballet:2020hze},
\begin{equation}
  \frac{d N}{d E}
=
  N_0 \left( \frac E {E_{0}} \right)^{-\alpha-\beta \log (E / E_{0})},
\label{Background}
\end{equation}
where $N_0$ is normalization, $E_0$ a scale parameter,
$\alpha$ the spectral slope at $E_0$, and $\beta$ the
curvature of the spectrum. Since $E_0$ does not vary
much, its value is fixed to 4074\,MeV in our fit while
the other three can freely adjust.

For a given DM mass $m_N$ and mass difference $\delta$,
the Yukawa coupling constant $y$ is fixed by the DM relic
density as explained earlier. This means that
the predicted photon flux around the SMBH is also 
determined for every parameter space point in
\gfig{Limit_Contour}. We fit the Fermi-LAT data with
the following $\chi^2$ function, 
\begin{equation}
\chi^2 
\equiv
\sum_i^{\text {bins }}
\left(\frac{
N_i^{\mathrm{data}}
- N_i^{\mathrm{bkg}}
- N_i^{\mathrm{NP}}}
{\sigma_i}\right)^2.
\label{x2function}
\end{equation}
In each bin, $N_i^{\mathrm{data}}$ is the data point, 
$N_i^{\mathrm{bkg}}$ the background,
$N_i^{\mathrm{NP}}$ the expected events due to new physics (NP) contribution, and 
$\sigma_i$ the error bar for each data point.

\gfig{Limit_Contour} shows the 95\% C.L. limit
as red lines and the region above are excluded. Since 
the realistic background is not determined and the 
parameters in \geqn{Background} are free,
we do the fit in two possible ways.
We first fix the background to its best-fit values obtained
from the background-only hypothesis with
$N_0 = 2.519 \times 10^{-12}$, $\alpha = 2.489$,
and $\beta = 0.1553$. In this case, the background 
model can fit data very well and consequently the resulting
limit on the signal parameters (red solid line) are quite
stringent. The allowed 
DM mass is below $\lesssim 60$\,GeV and the mass difference
$\delta$ cannot be larger than $\mathcal O(10^{-5})$.
For comparison, we also fit the data by changing both the
prediction and background parameters. The corresponding
result (red dashed line) allows much larger parameter space
with a mass difference as large as $10^{-3}$. Both methods
require the mass difference within a quite narrow range
at the leave of $\mathcal O(10^{-5})$ and $\mathcal O(10^{-3})$,
respectively for DM mass $m_N \gtrsim 30$\,GeV.

{\bf Conclusion and Discussion} --
In this paper, we propose a model with 
dark sector including two Majorana fermions 
$N$ and $\chi$, as well as a Higgs-portal scalar particle $\phi$.  
Among them, the right-handed neutrino $N$ is the lightest
one and hence can serve as DM.
Its freeze-out production in the early Universe is through
the forbidden annihilation channel
$N +  N \leftrightarrow \phi + \phi$. Although this annihilation
channel is nowadays forbidden, it can reopen 
around a strong gravitation source. Thus, the 
SMBH located at the galaxy center can accelerate 
$N$ to make it annihilate into its heavier partner $\phi$.
The subsequent decay of $\phi$ to SM particles 
can bring detectable photon flux. The current 
Fermi-LAT data requires the right-handed neutrino mass
$m_N$ smaller than $100$\,GeV, and the mass difference
$\mathcal O(10^{-5}) \lesssim \delta \lesssim \mathcal O(10^{-3})$. 

Such a light right-handed neutrino can be tested at future
lepton colliders such as CEPC \cite{CEPCStudyGroup:2018ghi},
ILC \cite{ILC:2013jhg}, FCC-ee \cite{FCC:2018evy}, and CLIC \cite{Linssen:2012hp}.
If the masses of $N$ and $\chi$ satisfy $m_\chi + m_N < m_h$, 
the SM Higgs particle can decay invisibly,
$h \rightarrow \chi + N$. For illustration, the projected
sensitivity $Br_{\mathrm{inv}}^{\mathrm{BSM}} < 0.3 \%$
at CEPC translates to $\lambda \lesssim 5 \times 10^{-2}$.
 With our parameter assignments,
$m_\phi > m_N$ and $m_\chi > 2 (m_N + m_\phi)$, this Higgs
invisible decay constraint applies for $m_N < 25$\,GeV
as shown with verticle black dashed line in \gfig{Limit_Contour}.
Besides, the dark sector particle $\phi$ can also be
directly produced at future Higgs factories through
Higgsstrahlung-like process $e^+ + e^- \rightarrow Z + \phi$
\cite{Ko:2016xwd,Heng:2017jgh,Liu:2017lpo,Kamon:2017yfx,Azevedo:2018oxv,Grzadkowski:2020frj,Haghighat:2022qyh}.

In addition to the current model assignments discussed
in this paper, there is another possibility of removing 
the dark sector particle $\chi$ and assigning the right-handed 
neutrino DM $N$ play its role. Then the forbidden
annihilation $N N \rightarrow \phi \phi$ arises from a
much simpler flavor-conserving Yukawa coupling $y \phi NN$.
One side effect of this model is introducing DM self-scattering 
$N N \leftrightarrow N N$ by exchanging a scalar mediator
$\phi$. Consequently, DM has much a smaller density profile
in the spike region. This leads to a weaker constraint on 
the annihilation cross section \cite{FuturePaper}.

\section*{Acknowledgements}
SFG is supported by the Double First Class start-up fund
(WF220442604) and
National Natural Science Foundation of China (No.\,12090064),
and Chinese Academy of Sciences Center for Excellence in Particle
Physics (CCEPP).
T. T. Y. is supported by the China Grant for
Talent Scientific Start-Up Project and by Natural Science
Foundation of China (NSFC) under grant No.\,12175134,
JSPS Grant-in-Aid for Scientific Research
Grants No.\,19H05810, 
and World Premier International Research Center
Initiative (WPI Initiative), MEXT, Japan.
Both SFG and T. T. Y.
are affiliated members of Kavli IPMU, University of Tokyo.

\providecommand{\href}[2]{#2}\begingroup\raggedright\endgroup

\vspace{15mm}
\end{document}